\documentclass[reprint,aps,amsmath,amssymb,twoside,prd,showkeys,superscriptaddress]{revtex4-1}

\pdfoutput=1

\usepackage{verbatim}
\usepackage{comment}
\usepackage{graphicx}
\usepackage[T1]{fontenc}

\usepackage{epsfig}
\usepackage{bm}
\usepackage{amssymb}
\usepackage{float}
\usepackage{amsmath}
\usepackage{subfigure}
\usepackage{dcolumn}
\usepackage{cancel}
\usepackage[colorlinks]{hyperref}
\usepackage[usenames,dvipsnames]{color}
\hypersetup{
     breaklinks=true,
    pdfstartview={FitH},    
    colorlinks=true,       
    linkcolor=blue,          
    citecolor=red,        
    filecolor=magenta,      
    urlcolor=blue,           
    anchorcolor=green,      
    linktocpage=true
}

\def\doi{http://doi.org}

\newcommand{\HCd}{\mathcal{H}}

\def\HCdt0{\tilde{\HCd}_{0}}

\newcommand{\afffias}{Frankfurt Institute for Advanced Studies (FIAS), 
Ruth-Moufang-Strasse~1, 60438 Frankfurt am Main, Germany}
\newcommand{\affbgu}{Physics Department, Ben-Gurion University of the Negev, 
Beer-Sheva 
84105, Israel}
\newcommand{\affbahamas}{Bahamas Advanced Study Institute and Conferences, 4A 
Ocean 
Heights, Hill View Circle, Stella Maris, Long Island, The Bahamas}

\newcommand{\affgreece}{Department of Physics, National Technical University of 
Athens, 
Zografou Campus GR 157 73, Athens, Greece}
\newcommand{\affgreecebb}{National Observatory of Athens, Lofos Nymfon, 11852 
Athens, 
Greece}

\newcommand{\affchaina}{Department of Astronomy, School of Physical Sciences, 
University 
of Science
and Technology of China, Hefei 230026, P.R. China}

\begin{document}

\title{The Scale Factor Potential Approach to Inflation}
 
\author{David Benisty}
\email{benidav@post.bgu.ac.il}
\affiliation{\affbgu}\affiliation{\afffias}

\author{Eduardo I. Guendelman}
\email{guendel@bgu.ac.il}
\affiliation{\affbgu}\affiliation{\afffias}\affiliation{\affbahamas}
\author{Emmanuel N. Saridakis}
\email{msaridak@phys.uoa.gr}
\affiliation{\affgreece}\affiliation{\affgreecebb}\affiliation{\affchaina}

\begin{abstract}
We propose a new approach to investigate inflation in a model-independent way, 
and in particular to elaborate the involved observables, through the 
introduction of the ``scale factor potential''. Through its use one can 
immediately determine the inflation end, which corresponds to its first (and 
global) minimum. Additionally, we express the inflationary observables in terms 
of its logarithm, using as independent variable the  e-folding number. As an example, we construct  a new class of scalar potentials 
that can lead to the desired spectral index and 
tensor-to-scalar ratio, namely $n_s \approx 0.965$ and $r \sim 10^{-4}$ for 60 
$e$-folds, in agreement with observations.
\end{abstract}
 

\maketitle

\section{Introduction} 

The inflationary paradigm is considered as a necessary part of the standard 
model of 
cosmology, since it provides   the solution to the fundamental puzzles of the 
old Big 
Bang theory, such as the horizon, the flatness, and the monopole problems
\cite{Starobinsky:1979ty,Kazanas:1980tx,
Starobinsky:1980te,Guth:1980zm,Linde:1981mu,
Albrecht:1982wi,Barrow:1983rx,Blau:1986cw}. It can be 
achieved through various mechanisms, for instance through the introduction of 
primordial scalar field(s) 
\cite{Barrow:2016qkh,Barrow:2016wiy,Olive:1989nu,Linde:1993cn,Liddle:1994dx,
Lidsey:1995np,CervantesCota:1995tz,Berera:1995ie,ArmendarizPicon:1999rj,
Kanti:1999ie,Garriga:1999vw,Gordon:2000hv,Bassett:2005xm,Chen:2009zp,
Germani:2010gm,Kobayashi:2010cm,Feng:2010ya,Burrage:2010cu,Kobayashi:2011nu,
Ohashi:2012wf,Hossain:2014xha,Hossain:2014zma,Cai:2014uka,Geng:2015fla,
Kamali:2016frd,Geng:2017mic,Benisty:2017lmt,Dalianis:2018frf,Dalianis:2019asr,
Benisty:2020vvm,Benisty:2019pxb,Benisty:2019bmi,Benisty:2019tno,
Staicova:2019ksr,Staicova:2018ggf,Staicova:2018bdy,Guendelman:1999rj,Guendelman:2014bva,Qiu:2020qsq}, or 
through correction terms into the modified gravitational action 
\cite{Dvali:1998pa,Kawasaki:2000yn,Bojowald:2002nz,Nojiri:2003ft,Kachru:2003sx,
Nojiri:2005pu,Ferraro:2006jd,Cognola:2007zu,Cai:2010kp,Ashtekar:2011rm,
Qiu:2011zr,Briscese:2012ys,Ellis:2013xoa,Basilakos:2013xpa,Sebastiani:2013eqa,
Baumann:2014nda,Dalianis:2015fpa,Kanti:2015pda,DeLaurentis:2015fea,
Basilakos:2015yoa,Bonanno:2015fga,Koshelev:2016xqb,Bamba:2016wjm,
Motohashi:2017vdc,Oikonomou:2017ppp,Benisty:2018fja,Antoniadis:2018ywb,
Karam:2019dlv,Nojiri:2019kkp,Benisty:2019jqz,Mukhanov:2013tua}.

Additionally, inflation was proved crucial in  
providing a framework for the generation of primordial density perturbations 
\cite{Mukhanov:1981xt,Guth:1982ec}. Since these perturbations affect  the 
Cosmic Background Radiation (CMB), the inflationary effect on observations can 
be 
investigated through the prediction for the  scalar spectral index of the 
curvature 
perturbations and its running, for the tensor spectral index, and for the 
tensor-to-scalar ratio.  

The standard approach to calculate the above inflation related observables, is 
by 
performing a detailed perturbation analysis. Nevertheless, the procedure can be 
simplified if one imposes the slow-roll approximation and introduces the 
slow-roll 
parameters  \cite{Martin:2013tda}, either in the case where inflation is driven 
by a 
scalar field and its potential, or in the case where inflation arises through 
gravitational modification.

In the present work we propose a new approach to investigate inflation, and in 
particular the involved observables, through the introduction of the ``scale 
factor 
potential''. This scale factor potential   is defined by   demanding it to be
opposite to the ``kinetic energy''   of the scale factor  in order for them to 
add to 
zero. As we will see, it is very useful in studying inflation for every 
underlying 
theory, since through its use one can immediately determine the inflation end, 
namely at 
its minimum, as well as he can calculate the various inflationary observables.

The plan of the work is as follows: In section \ref{modelsc} we introduce the 
concept of 
the scale-factor potential. In section \ref{Applicationsc} we apply it in order 
to 
investigate inflation in general, and using it we propose a new inflationary 
scalar-field 
potential that can generate a spectral index and a tensor-to-scalar ratio in 
agreement 
with observations. Finally, in section \ref{Conclusionssc} we summarize our 
results.

\section{Scale factor potential}
\label{modelsc}

In this section we introduce the concept of ``scale factor potential'', which is 
a 
mathematical tool that proves very useful in performing inflationary 
calculations.
We focus on the usual case of a homogeneous and isotropic cosmology with the
Friedmann-Robertson-Walker (FRW) metric
\begin{eqnarray}
{\rm d}s^2 = -{\rm d}t^2 + a^2 (t) \left[\frac{{\rm d}r^2}{1-Kr^2} + r^2 \left(d 
\theta^2 
+ \sin^2 \theta d \phi^2\right)\right],
\end{eqnarray}
where $a(t)$  is the scale factor  and $K$ determines spatial curvature, with 
value of 
$0$ for a spatially flat universe.

\begin{figure}[ht]
 	\centering
\includegraphics[width=0.49\textwidth]{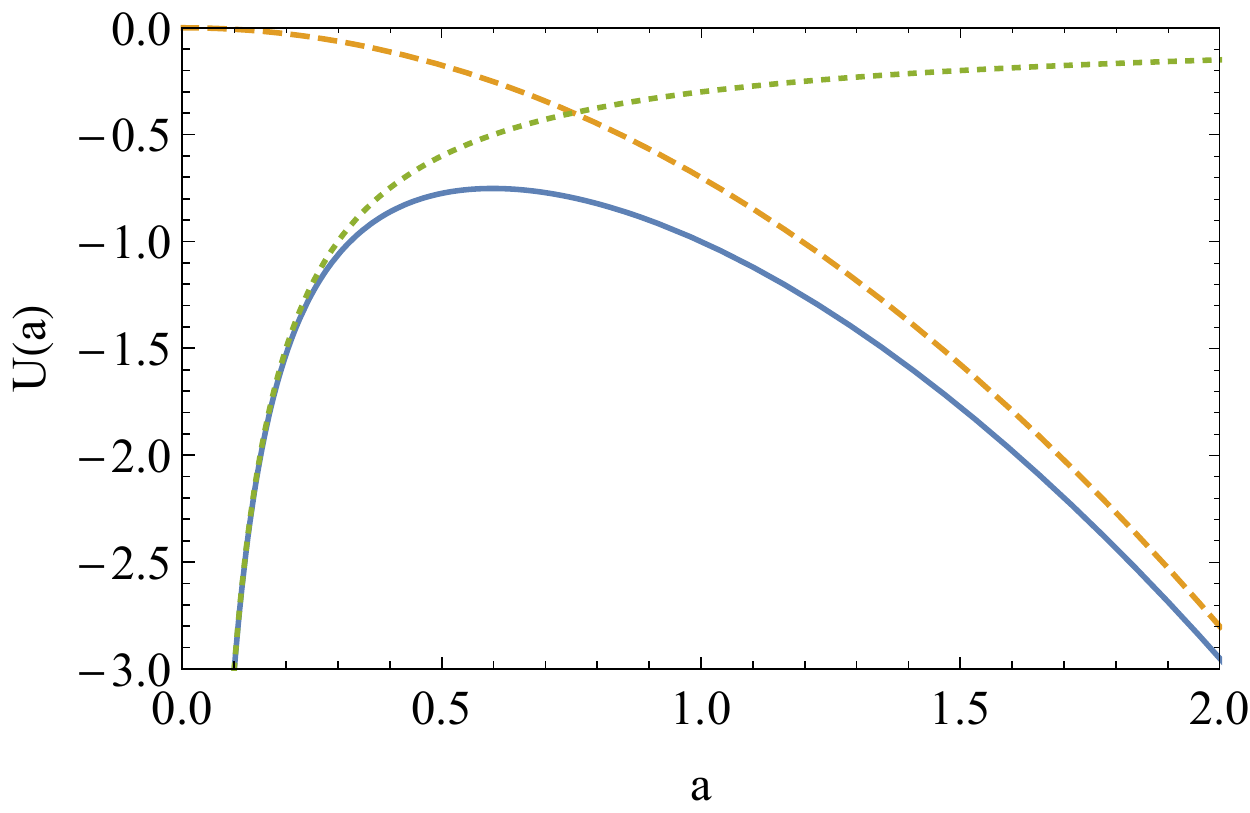}
\caption{{\it{The scale factor potential $U(a)$ for the case of a flat de Sitter 
Universe 
(orange-dashed curve), i.e with $\Omega_\Lambda^{(0)}=1$, for the case of a flat
matter-dominated universe (green-dotted curve), i.e with 
$\Omega_m^{(0)}=1$, and for $\Lambda$CDM paradigm (blue-solid curve), with 
$\Omega_\Lambda^{(0)}=0.7$, $\Omega_m^{(0)}=0.3$, in units where $H_0=1$. }}}
\label{fig1}
\end{figure}

The scale factor potential $U(a)$  is defined by   demanding it to opposite to 
the 
``kinetic energy''   of the scale factor, i.e. $\dot{a}^2$, in order for them to 
add to 
zero, namely:
\begin{equation}\label{potential}
  \dot{a}^2 + U\left( a\right) = 0,
\end{equation}
and hence it has dimensions of inverse length square. 
In order to provide a more illustrating picture, let's consider the general 
Friedmann 
equation in the case of $\Lambda$CDM paradigm, namely 
\begin{equation}
H^2 + \frac{K}{a^2} = \frac{8\pi G}{3} \left(\rho_m+\rho_r+\rho_\Lambda
\right),\label{FR1} 
\end{equation}
where $\rho_m$,$\rho_r$,$\rho_\Lambda$ correspond respectively to the energy 
density of 
 matter, radiation and cosmological constant, and $G$ is the Newton's constant. 
Hence, 
in this case the corresponding scale factor potential will be 
\begin{equation}
U(a) = -  a^2 H_0^2  \left[
\Omega_\Lambda^{(0)} + \frac{\Omega_K^{(0)}}{a^2} + 
\frac{\Omega_m^{(0)}}{a^3} + \frac{\Omega_r^{(0)}}{a^4}
\right],
\end{equation}
where $\Omega_i^{(0)}$ are the values of the density parameters $\Omega_i=8\pi 
G\rho_i/3H^2$ at the present scale factor $a_0=1$, and $H_0$ is the present 
Hubble 
parameter (we have defined $\rho_K\equiv -3K/(8\pi G a^2)$). In Fig. \ref{fig1} 
we depict 
$U(a)$ for the case where the Universe contains only the cosmological constant 
(de Sitter 
Universe), for the case of a matter-dominated Universe, and for the standard 
$\Lambda$CDM 
scenario.

\section{Application to Inflation}
\label{Applicationsc}

In this section we investigate the inflation realization using the scale factor 
potential introduced above. Let us first start by the description of the basic 
de Sitter 
evolution. One can immediately see that in such exponential expansion of the 
form 
$a(t) = a_i e^{H_{dS} (t-t_i)}  $ the scale factor potential (\ref{potential}) 
is just an 
inverse parabola, namely $U(a) = - H_{dS}^2a^2$,  whose shape is determined by 
the de 
Sitter Hubble  parameter value $H_{dS}$. Hence, we deduce that in any 
physically 
interesting inflationary scenario, the scale factor potential will start from an 
inverse 
parabola at small scale factors, and then as the universe proceeds towards the 
inflationary exit $U(a)$ will deviate accordingly.

The important issue in a successful inflationary realization is the calculation 
of 
various inflation-related observables, such as the scalar spectral index of the 
curvature 
perturbations $n_\mathrm{s}$, its running $\alpha_\mathrm{s} \equiv d 
n_\mathrm{s}/d 
\ln k$, where $k$ is the absolute value of the wave number $\Vec{k}$,
the tensor spectral index $n_\mathrm{T}$ and the tensor-to-scalar ratio $r$. 
These 
quantities are determined  by observational data very accurately, and hence 
confrontation 
can constrain of exclude the studied scenarios.

In general, the calculation of the above observables demands a detailed  
perturbation 
analysis. Nevertheless, one can obtain approximate expressions by imposing the 
slow-roll assumptions, under which all  inflationary information is encoded in 
the 
slow-roll parameters. In particular, one first introduces  
\cite{Martin:2013tda} 
\begin{equation}\label{slowRoll}
  \epsilon_{n+1} = \frac{d}{dN} \log 
|\epsilon_n|,
\end{equation}
where $\epsilon_0\equiv H_{i}/H$ and $N\equiv\ln(a/a_{i})$ is the e-folding 
number, with  $a_{i}$   the scale factor at the beginning of inflation, $H_{i}$ the corresponding Hubble parameter, and $n$ a positive integer. As usual inflation ends at a scale factor $a_f$  
where $\ddot{a} = 0$, i.e. where $\epsilon_1(a_f)=1$, and the slow-roll 
approximation breaks down. Finally, in terms 
of the 
first three $\epsilon_n$, which are easily found to be 
\begin{eqnarray}
\label{epsVbb}
&&\!\!\!\!\!\!\!\!\!\!\!\!\!\!
\epsilon_1\equiv-\frac{\dot{H}}{H^2}, 
\\
&&\!\!\!\!\!\!\!\!\!\!\!\!\!\!
\epsilon_2 \equiv  \frac{\ddot{H}}{H\dot{H}}-\frac{2\dot{H}}{H^2},
\label{etaVbb}\\
&&\!\!\!\!\!\!\!\!\!\!\!\!\!\!
\epsilon_3 \equiv
\left(\ddot{H}H-2\dot{H}^2\right)^{-1}
\nonumber\\
&&\!\!\!\!
\cdot\!\left[\frac{H\dot{H}\dddot{H}-\ddot{H}(\dot{H}
^2+H\ddot{H}) } { H\dot { H } }-\frac{2\dot{H}}{H^2}(H\ddot{H}-2\dot{H}^2)
\right]\!,
\label{xiVbb}
\end{eqnarray}
the inflationary observables are expressed as \cite{Martin:2013tda}
 \begin{eqnarray}
 r &\approx&16\epsilon_1 ,
 \label{eps111bb}\\
n_\mathrm{s} &\approx& 1-2\epsilon_1-\epsilon_2 ,
\label{eps2222bb} \\
\alpha_\mathrm{s} &\approx& -2 \epsilon_1\epsilon_2-\epsilon_2\epsilon_3  ,
\label{eps333bb}\\
 n_\mathrm{T} &\approx& -2\epsilon_1 ,
\label{eps444bb}
\end{eqnarray}
where all quantities are calculated at $a_{i}$.

Let us now see how the above approach is simplified with the use of the scale 
factor 
potential $U(a)$. In particular, using the definition (\ref{potential}) we can 
immediately express the slow-roll parameters above as:
 \begin{eqnarray}
 \label{epsilon1}
&&
\!\!\!\!\!\!\!\!\!\!\!
\epsilon_1   = 1-\frac{a U' }{2 U },
\\
&&
\label{epsilon2}
\!\!\!\!\!\!\!\!\!\!\!
\epsilon_2   = \frac{a 
\left\{a U'^2-U  \left[a 
U'' +U' \right]\right\}    }
{U  \left(2 U -a U' \right)    },
\\
&&
\label{epsilon3}
\!\!\!\!\!\!\!\!\!\!\!
\epsilon_3   = \left\{U \left(2 
U-a U'\right) \left[U \left(a U''+U'\right)-a 
U'^2\right]\right\}^{-1}\nonumber\\
&&
\
\cdot
\Big\{-a^3 U'^4+a^2 U U'^2 \left(a 
U''+5 U'\right)\nonumber\\
&& \ \ \ \ \,
-a U^2 \left[-a^2 U''^2+a U' \left(a U'''+7 U''\right)+6 
U'^2\right]\nonumber\\
&& \ \ \ \ \,
+2 U^3 \left[a \left(a U'''+3 U''\right)+U'\right]\Big\}
,
\end{eqnarray}
where primes denote derivatives with respect to $a$. The end of inflation is 
obtained 
when   $\epsilon_1 (a_f) = 1$. Eq.  
(\ref{epsilon1}) with $\epsilon_1 = 1$ yields $U'(a_f)=0$.  Hence, we deduce 
that inflation ends at the minimum of the scale
factor potential (we know that it is minimum and not a maximum 
since as we mentioned the evolution in every inflationary model starts close to 
de Sitter i.e.  
to the inverse parabola $U(a) = - H_{dS}^2a^2$, thus it starts from a maximum 
of 
 $U(a)$. The simplicity of the condition  $U'(a_f)=0$ reveals the 
advantage of the use of $U(a)$). 
This feature will become useful later on. Finally, by inserting relations  
(\ref{epsilon1})-(\ref{epsilon3}) calculated at $a_i$ into 
(\ref{eps111bb})-(\ref{eps444bb}) we obtain the inflationary observables.

Since the e-folding number is defined as the logarithm of the scale factor, 
namely  
$N\equiv\ln(a/a_{i})$, we can introduce the logarithm of 
the scale factor potential as
\begin{equation}\label{logPam}
  P = -\ln \left[\frac{U(a)}{U(a_i)} \right].
\end{equation}
 Using these variables the  Hubble function is expressed in terms of the 
e-folding number 
as
\begin{equation}\label{HubbleCon}
H(N) = H(0) \,  \exp \left[-N -\frac{1}{2}P(N) \right],
\end{equation}
which proves to be   very useful   since it is straightforwardly relates $H$ 
with $N$, 
i.e. to the variable which determines the duration of a successful inflation (a 
successful inflation needs $\mathcal{N}_f\sim50-70$). Finally, inserting these 
variables into 
(\ref{epsilon1})-(\ref{epsilon3}) we express the slow-roll parameters is a 
simple way as  
(\ref{slowRoll}):
 \begin{eqnarray}
&&\label{epsilon1b}
\epsilon_1 = 1 + \frac{1}{2} P'\left(N\right),
\\\label{epsilon2b}
&&\epsilon_2 = \frac{P''\left( N \right)}{P'\left( N \right)+2},
\\
&&\epsilon_3 = \frac{P'''(N)}{P''(N)}-\frac{P''\left( N \right)}{P'\left( N 
\right)+2}.
\label{epsilon3b}
\end{eqnarray}
Since inflation ends when $\epsilon_1(\mathcal{N}_f)=1$, from (\ref{epsilon1b}) 
we deduce that this 
happens at $P'(\mathcal{N}_f) = 0 $, i.e at the minimum of $P$, which was 
expected since as we 
mentioned above inflation ends at the minimum of $U$.  

Inserting relations  
(\ref{epsilon1b})-(\ref{epsilon3b}) calculated at the beginning of inflation, 
i.e. at  
$N=0$, into 
(\ref{eps111bb})-(\ref{eps444bb}) we obtain the inflationary observables. In 
particular, 
doing so we find:
 \begin{eqnarray}
 &&r \approx  16 + 8 P'\left(0\right),
 \label{eps111bbcd}\\
&&  n_\mathrm{s} \approx 
 -1-   P'\left(0\right)-\frac{P''\left( 0 \right)}{P'\left( 0 \right)+2},
\label{eps2222bbcd} \\
&&\alpha_\mathrm{s} \approx -P''(0)-\frac{P'''(0)}{P'(0)+2}+\left[
\frac{P''(0)}{P'(0)+2}\right]^2  ,
\label{eps333bbcd}\\
 &&n_\mathrm{T} \approx -2 -  P'\left(0\right).
\label{eps444bbcd}
\end{eqnarray}
Hence, as we can see, the initial values for $P$ and its derivatives, i.e. of 
the scale 
factor potential and its derivatives, are the crucial ones in determining the 
value of 
the inflationary observables.
In the slow-roll approximation in the beginning of 
inflation we have  $\epsilon_n \ll 
1$, which using expressions (\ref{epsilon1b})-(\ref{epsilon3b}) lead to
 \begin{eqnarray}
&&-2 \lesssim P'\left(0\right) \ll 0
\nonumber\\
&&
0 \lesssim P''(0) \ll P'(0) + 2.
\label{eq:Pcon}
\end{eqnarray}

We proceed by exploring the properties of the logarithm of the scale factor 
potential 
$P(N)$ in order to obtain inflationary observables, and in particular spectral 
index $ 
n_\mathrm{s}$  and tensor-to-scalar ratio $r$, in agreement with observations. 
From 
(\ref{eps111bbcd}),(\ref{eps2222bbcd}) we   acquire  
 \begin{eqnarray}\label{rel1}
&&P'(0) = \frac{r}{8} - 2
\\
&&\label{rel2}
P''(0) = \frac{r}{64} \left[8(1- n_\mathrm{s}) - r \right].
\end{eqnarray}
 Hence, we need to introduce a parametrization for $P(N)$ that could incorporate 
these.
From the definition (\ref{HubbleCon}) we find that  the pure de Sitter solution
 gives  $P_{dS} = - 2 N$, and 
thus  $P_{dS} (0 ) = 0$, $P_{dS}' (0) = -2$,  $P_{dS}'' (0)= 0$, which 
corresponds to the 
inverse parabola behavior of the scale factor potential mentioned above. Since 
the bulk 
of inflation corresponds to an exponential expansion, a good parametrization for 
$P(N)$ 
should be a suitable deviation from this de Sitter form.

The above scale factor potential formalism is of general applicability in any 
inflation realization, whether this is driven by a scalar field, or it arises 
effectively from modified gravity, or from any other mechanism. In order to 
provide a more transparent picture let us consider as an example the well-known 
Starobinsky inflation 
\cite{Starobinsky:1979ty,Barrow:1983rx,Barrow:1988xh,Barrow:1988xi}. 
This scenario arises from a quadratic $f(R)$ gravity of the form  $f(R)= 
\frac{1}{16\pi G}R + \frac{1}{2M^2}R^2$, with $M$ a mass scale, which 
transformed in the Einstein frame is equivalent with a canonical scalar field 
$\phi$ moving in a potential \cite{Barrow:1988xh,Barrow:1988xi}: 
 \begin{equation}
V\left(\phi \right) =\frac{M^2}{32\pi G} 
\left( 1- e^{-\sqrt{16\pi  G /3} \phi } 
\right)^2.
\label{Starpot}
\end{equation}
\begin{figure}[t!]
 	\centering
\includegraphics[width=0.43\textwidth]{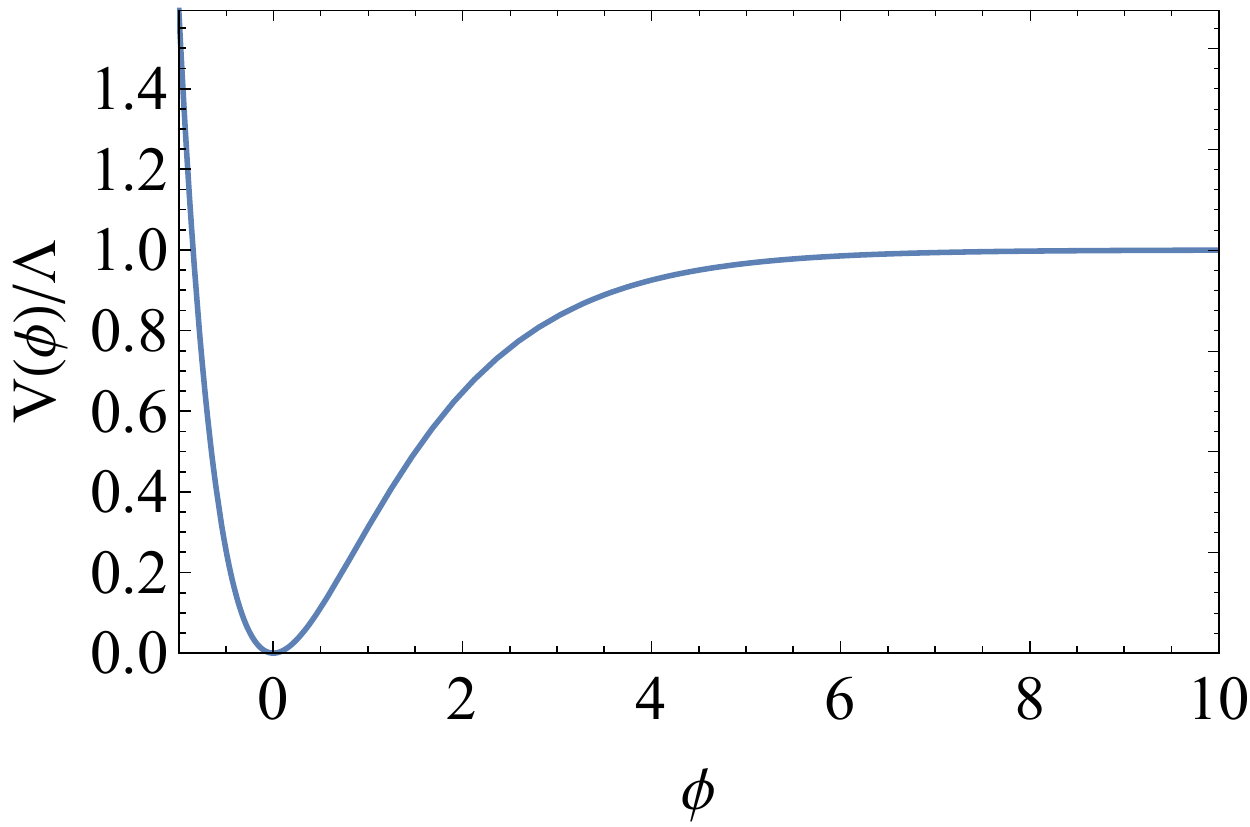}\\
\includegraphics[width=0.43\textwidth]{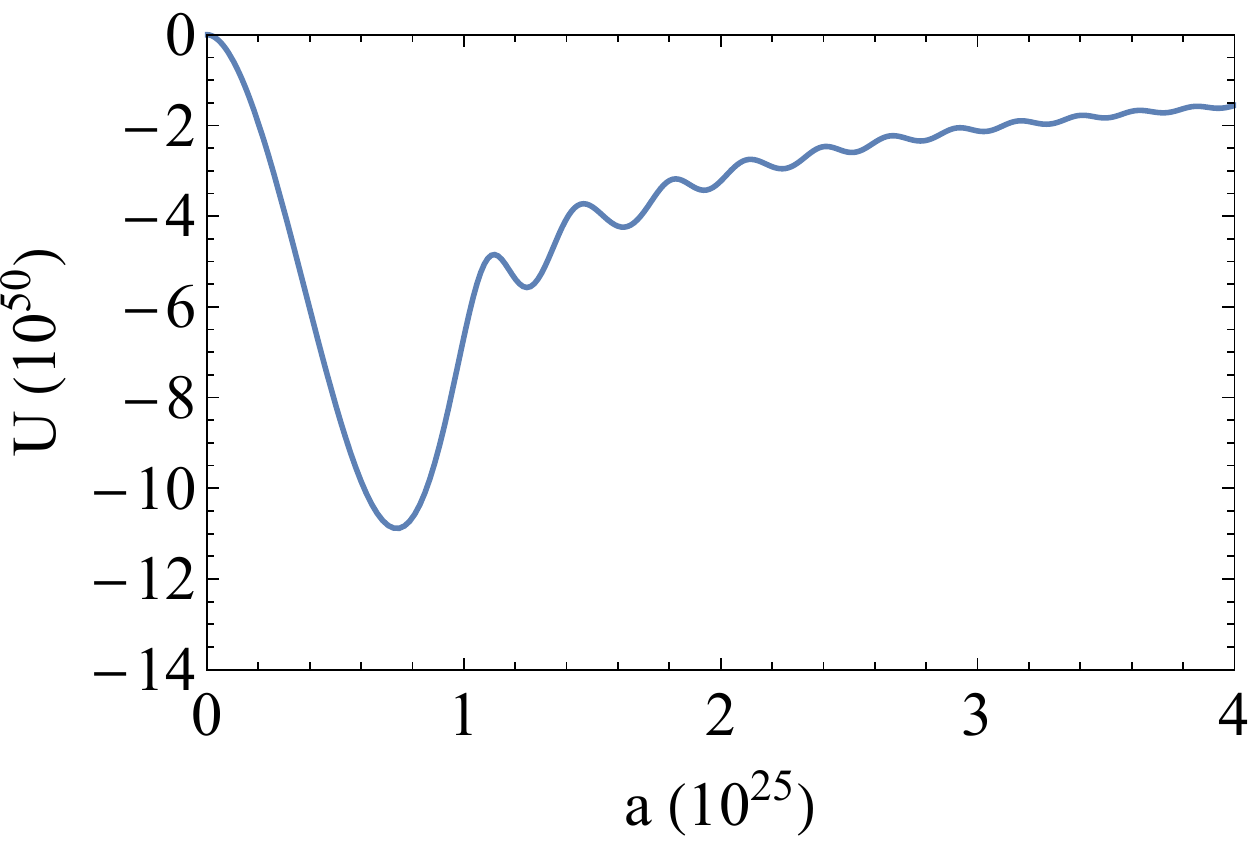}
\caption{{\it{
Upper  graph: The scalar field potential $V(\phi)$ in the Einstein frame of 
the Starobinsky inflation. Lower graph: The corresponding scale factor 
potential 
$U(a)$ 
as 
it is numerically reconstructed from the solution of Eqs. 
(\ref{Starpot})-(\ref{KleinG}).
Inflation ends at the global minimum of $U(a)$, and its subsequent 
oscillations  correspond to the scalar 
oscillations around the minimum of  $V(\phi)$ during the reheating 
phase.
We use units where $8\pi G=1$ and we set the scale factor at the beginning of 
inflation  
to $a_i=1$.
}} 
}
\label{fig3}
\end{figure}
The Friedmann equations are:
\begin{eqnarray}\label{FriedInf}
&&  H^2  = \frac{8\pi G}{3}\left[\frac{1}{2} \dot{\phi}^2 + V(\phi)\right], \\
&&
 \dot{H} = -4\pi G\dot{\phi}^2,
\end{eqnarray}
while the Klein Gordon equation for the scalar field is
\begin{equation}
\label{KleinG}
\ddot{\phi} + 3 H \dot{\phi} + V'(\phi) = 0.
\end{equation}
In the upper panel of Fig. \ref{fig3} we present the shape of the Starobinsky 
potential 
(\ref{Starpot}). On the lower panel we depict the corresponding scale factor 
potential as 
it is numerically reconstructed from the evolution of Eqs. 
(\ref{Starpot})-(\ref{KleinG}).
 As we observe, and as analyzed in detail above, the scale factor potential 
starts with an 
inverse parabola at the initial scale factors and inflation durates up to its 
first (and 
global) minimum. The subsequent oscillations of $U(a)$ correspond to the scalar 
oscillations around the minimum of the physical potential $V(\phi)$ during the 
reheating 
phase \cite{Kofman:1997yn}. Note the advantage that in the scale factor 
potential picture 
we know exactly the inflation end, namely at its minimum, while in the usual 
potential 
picture it is not straightforwardly determined when the slow roll finishes and 
inflation 
ends.

\section{Parameterization of the Potential}

  In this section we consider a specific example of the above formalism.
We apply the parametrization
\begin{equation}
P(N)=-2 \left(\frac{1}{N_0+N} +N\right),
\label{PolPot}
\end{equation}
with $N_0$ the model parameter. This form satisfies the condition 
(\ref{eq:Pcon}). Using that the end of inflation happens at $P'(\mathcal{N}_f) 
= 
0 $ it
gives:
\begin{equation}
\label{con111}
N_{f} + N_0 =  1.
\end{equation}
The corresponding slow-roll parameters  (\ref{epsilon1b})-(\ref{epsilon3b}) 
read:  
\begin{eqnarray}
&&\epsilon_1 = \frac{1}{(N_0+N)^{2}}\nonumber\\
&& \epsilon_2 = -\frac{2}{N_0+N}
\nonumber\\
&&\epsilon_3 = -\frac{1}{N_0+N}.
 \end{eqnarray}
  Moreover,   all   other slow-roll parameters are the same with $\epsilon_3$. 
As we can see, the advantage of the ansatz (\ref{PolPot}) is that all 
slow-roll parameters are small and therefore the initial state is by 
construction close to de Sitter solution.  The 
inflationary observables become
 \begin{equation}
 \label{pred1}
r = \frac{16}{(\mathcal{N}_f-1)^2} \end{equation}
\begin{equation}
\label{pred2}
n_s = \frac{(\mathcal{N}_f-4) \mathcal{N}_f+1}{(\mathcal{N}_f-1)^2}.
\end{equation}
Taking as an example the e-folding number as $\mathcal{N}_f=60$ we find that
\begin{equation}
r = 0.00459,\quad n_\mathrm{s} = 0.9655.
\label{eq:obvs}
\end{equation}
Eliminating $\mathcal{N}_f$ between (\ref{pred1})-(\ref{pred2}) gives 
\begin{equation}
r=-8 \left( n_\mathrm{s}-2+\sqrt{3-2 n_\mathrm{s}}\right),
\label{observ3}
\end{equation}
which is a very useful expression since it allows for a direct comparison with 
observations.
 \begin{figure}[ht]
 	\centering
\includegraphics[width=0.51\textwidth]{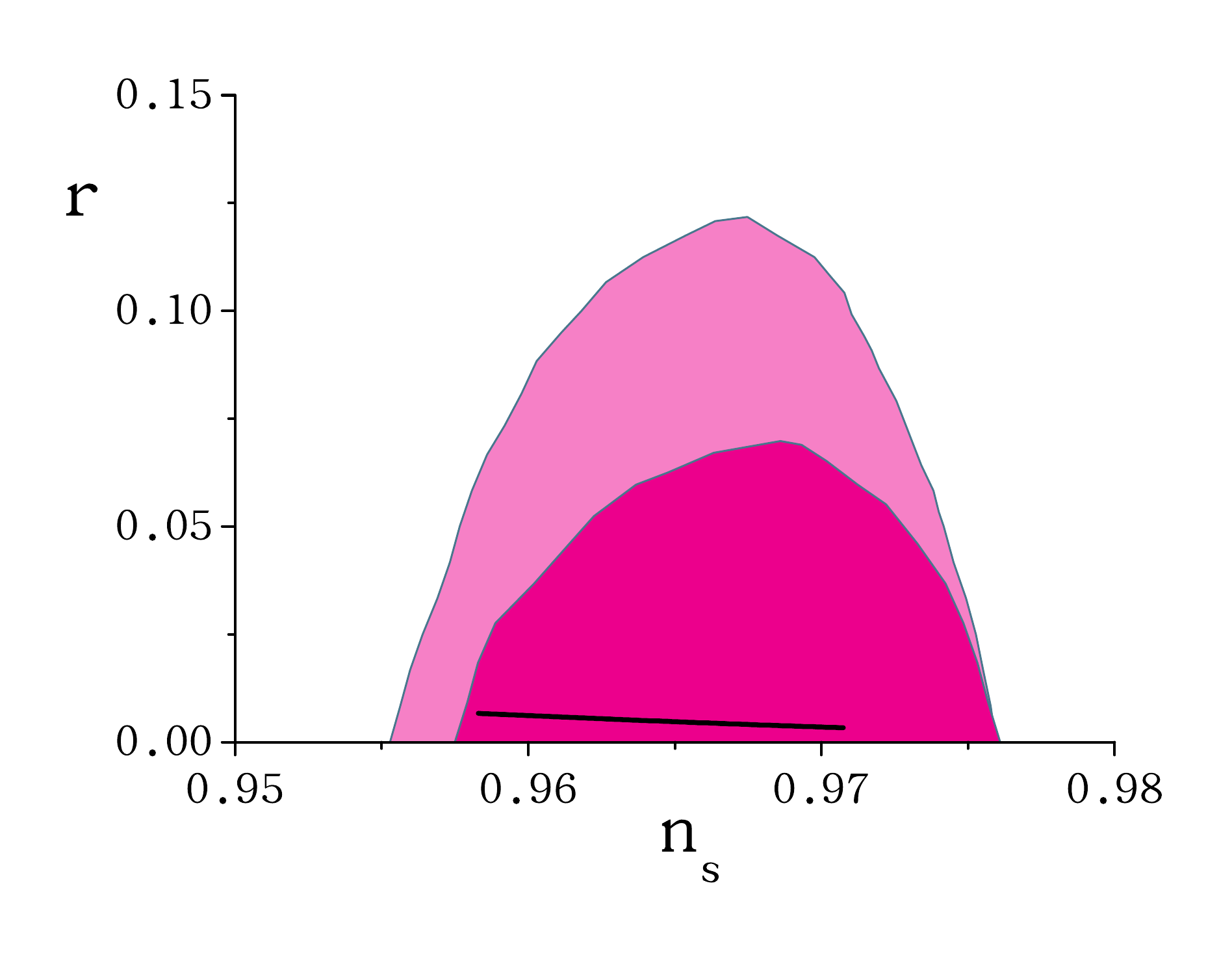}
\caption{{\it{1$\sigma$ (magenta) and 2$\sigma$ (light magenta) contours for 
Planck 2018
results (Planck $+TT+lowP$)  \cite{Akrami:2018odb}, on 
$n_{\mathrm{s}}-r$ plane.
Additionally, we present the predictions of the
scale factor potential (\ref{PolPot}) with $50\leq \mathcal{N}_f\leq 70$, 
according to 
(\ref{observ3}) ($n_{\mathrm{s}}=0.9584$ for  $\mathcal{N}_f=50$ and 
$n_{\mathrm{s}}=0.9706$ for  $\mathcal{N}_f=70$).}} 
}
\label{figobserv}
\end{figure}  
   
In Fig.~\ref{figobserv} we present the predictions of the scenario at hand in 
the $n_s -r$ plane, for e-folding numbers $\mathcal{N}_f$ varying between 50 
and 70, on top of the 1$\sigma$ and 
2$\sigma$ likelihood contours of the Planck 2018 results 
\cite{Akrami:2018odb,Aghanim:2018eyx}. As we 
can see, the  
agreement with observations is very efficient, and the predictions lie well 
within the 1$\sigma$ region. Moreover, in the future Euclid and SPHEREx 
missions and the BICEP3 experiment, are expected to provide better 
observational bounds
to test these predictions.

 We can now proceed in applying the scale factor potential approach in order to 
reconstruct a physical 
scalar-field potential that can generate the desirable inflationary observables. 
From the 
definition of the scale factor potential (\ref{potential}), as well as the 
Friedmann 
equation  (\ref{FriedInf}) that holds in every scalar-field inflation, we 
extract the 
following solutions:
  \begin{eqnarray}
&&
\phi(a) = - \int_{a_i}^{a} \frac{\sqrt{2 U(a )-a U'(a)}}{a \sqrt{U(a)}} \, 
da,\\
&&V(\phi(a)) = V_0-\frac{a\, U'(a)+4 U(a)}{2 a^2}.
\end{eqnarray}
Expressed in terms of the e-folding number $N$ and the logarithm of the scale 
factor 
potential $P(N)$ of (\ref{logPam}) the above solutions become:
 \begin{eqnarray}\label{conTran1}
&&
\phi(N) = - \int_{0}^{N} \sqrt{2 + P'(N)} \, dN,
\\
&&
\label{conTran2}
V(\phi(N)) = V_0 + e^{-P(N) - 2N} \left[2-\frac{1}{2}P'(N)\right].
\end{eqnarray}

\begin{figure}[ht]
\centering
\includegraphics[width=0.42\textwidth]{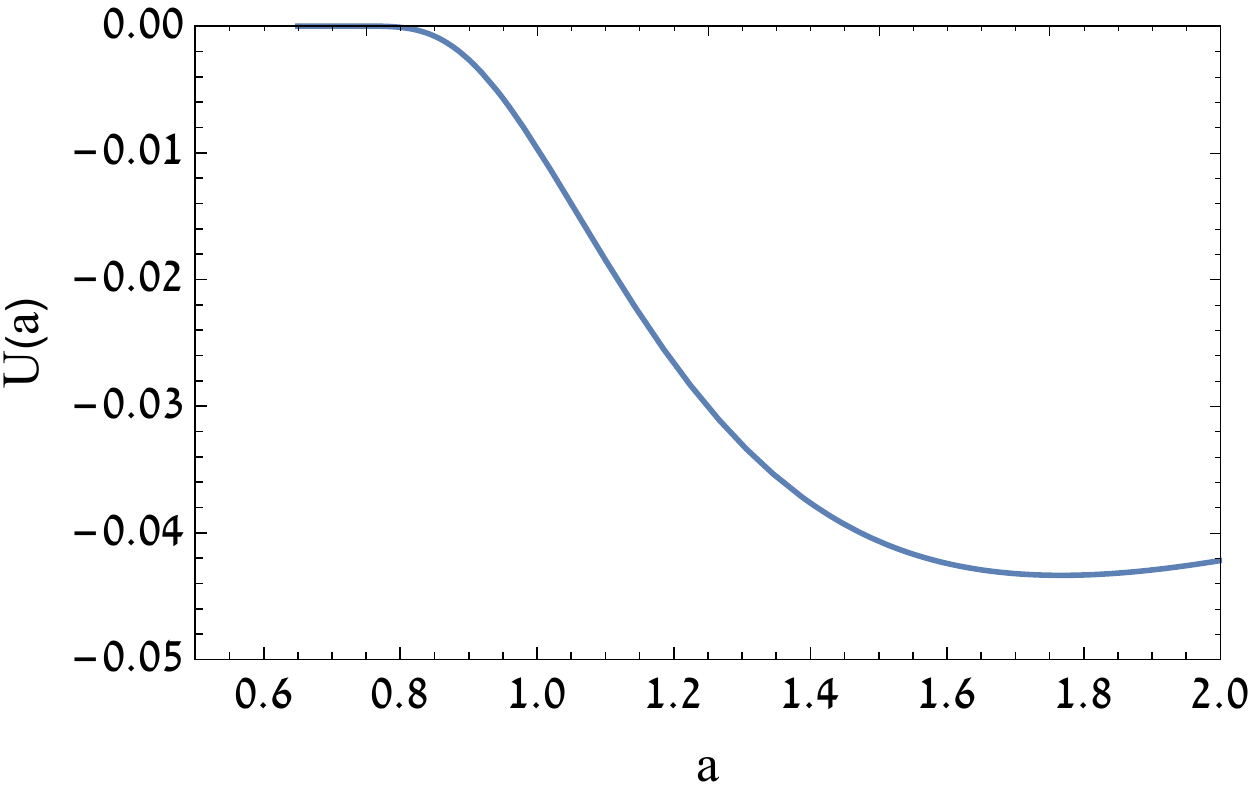}
\\
\includegraphics[width=0.43\textwidth]{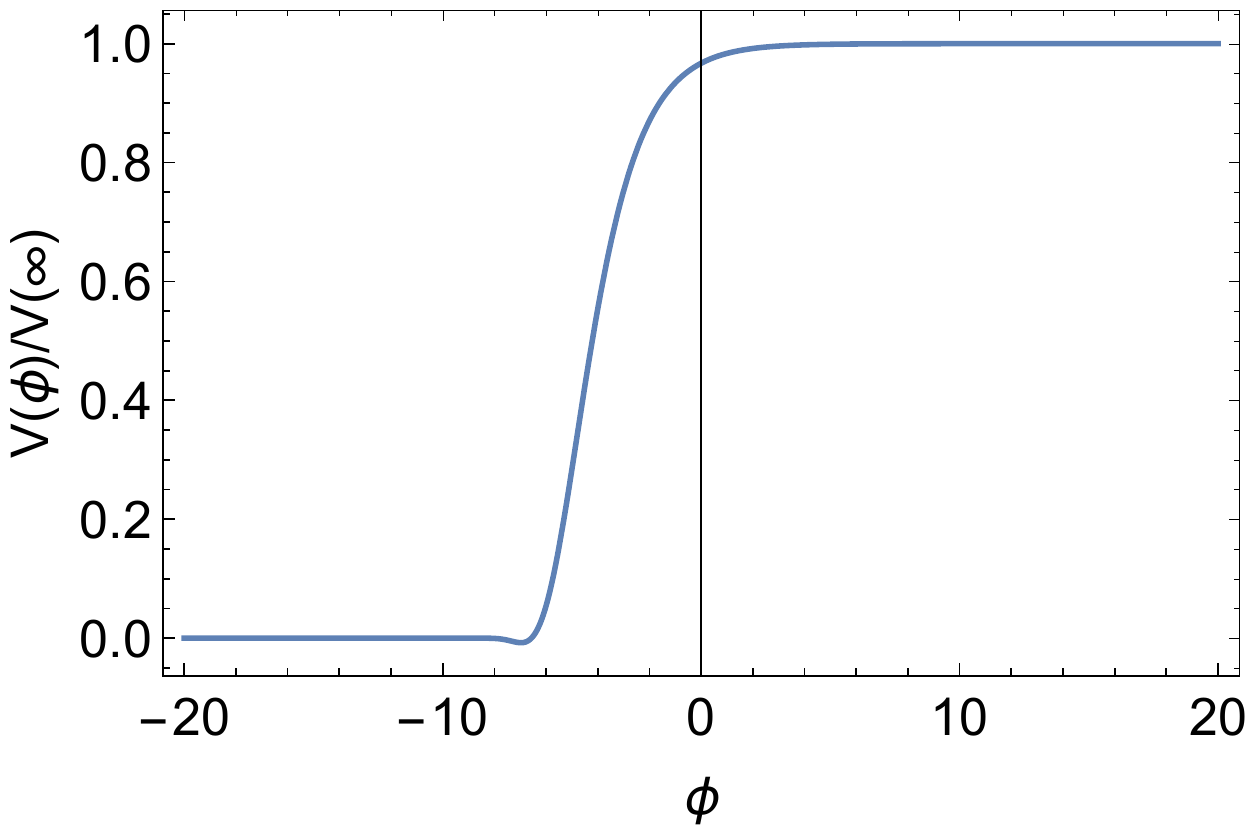}
\caption{{\it{
Upper  graph: The scale factor potential $U(a)$ of the parameterization
(\ref{PolPot}). 
 Lower graph: The corresponding scalar-field potential $V(\phi)$ of  
(\ref{FinPot}), normalized by its value at $\infty$. We use units where $8\pi 
G=1$ and we set 
the scale factor at the beginning of inflation to $a_i=1$. We consider    
e-folding number   $\mathcal{N}_f=60$, and thus (\ref{con111}) gives 
$N_0=-59$. Inflation 
ends at the global minimum of $U(a)$. }}}
\label{fig2}
\end{figure}

Let us   apply the above formalism in our specific  parametrization 
(\ref{PolPot}). Inserting it  into 
(\ref{conTran1})-(\ref{conTran2}) finally yields:
\begin{eqnarray}
 &&  
 \!\!\!\!\!\!\!\!\!\!\!\!\!\!\!\!
 \phi(N) = \sqrt{2} \log \left(\frac{N+ N_0}{N_0}\right)
\label{phiN1}
\end{eqnarray}
and
\begin{equation}
\begin{split}
V(\phi(N)) =H_0^2 e^{\frac{2}{N_0+N}} \left[3-\frac{1}{(N_0+N)^2}\right].     
\end{split}
\label{VphiN1}
\end{equation}
Expression (\ref{phiN1}) can be inversed, in order to find $N(\phi)$ and 
then 
 through 
insertion into (\ref{VphiN1}) to extract $V(\phi)$
analytically as 
\begin{eqnarray}
&&
 \!\!\!\!\!\!\!\!\!\!\!\!\!\!\!\!
 V(\phi) = H_0^2 e^{\frac{2 e^{-\frac{\phi } {\sqrt{2}}}}{N_0}-\sqrt{2} \phi } 
\left(3 N_0^2 e^{\sqrt{2} \phi }-1\right). 
\label{FinPot}
\end{eqnarray}
Hence, this potential   is the physical potential that leads to the 
observables depicted in Fig. \ref{figobserv}.
In Fig. \ref{fig2} we depict the scale factor potential $U(a)$ of the 
parameterization (\ref{PolPot}), as well as the corresponding 
scalar-field potential $V(\phi)$ of  
(\ref{FinPot}). The 
universe   begins with $\phi \gg1$ with a slow-roll 
behavior, and the scalar field moves towards the left.
The asymptotic 
values of the potential are:
\begin{equation}
V_{+\infty} = 3 H^2_{0}, \quad V_{-\infty} = 0,
\end{equation}
and thus   $3 H^2_{0}$ represents the energy   scale of the inflationary 
epoch.

We close this section by mentioning that the above potential reconstruction was 
just an 
example that arose from the consideration of the polynomial parametrization of 
$P(N)$ in 
(\ref{PolPot}). By imposing other parametrizations we can obtain, numerically 
or 
analytically, other potential forms that lead to the desired inflationary 
observables.
Such capabilities reveal the advantages of the approach at hand.

\section{Conclusions}
\label{Conclusionssc}
In this work we proposed a new approach to investigate inflation in a 
model-independent 
way, and in particular to elaborate the involved observables, through the 
introduction of the ``scale factor potential'' $U(a)$ . This     potential   is 
defined by   demanding it to be opposite to the ``kinetic energy''  of the scale 
factor  in order for them to add to zero. 

The scale factor potential is very useful in studying inflation for every 
underlying theory. Firstly, through its use one can immediately determine the 
inflation end, 
which corresponds to its first (and global) minimum, which is an advantage 
comparing to the usual potential picture, in which it is not straightforwardly 
determined when the 
slow roll finishes and inflation ends. The subsequent oscillations of $U(a)$ 
correspond to the scalar 
oscillations around the minimum of the physical potential  during the reheating 
phase. 

Additionally, we expressed the inflationary observables, such as the spectral 
index  and its running,  the  tensor-to-scalar ratio, and the tensor spectral 
index, in terms of the scale factor potential and its 
derivatives. Then we introduced the logarithm $P$ of $U$ and we used as 
independent 
variable the  e-folding number $N$, re-expressing the inflationary observables 
straightaway in terms of the initial values of $P$ and its derivatives. In this 
way, 
introducing parametrizations for $P(N)$ we were able to reconstruct $U$ that 
leads to 
the imposed inflationary observables.

We applied it in order to reconstruct a   physical 
scalar-field potential that can generate the desirable inflationary observables.
 Hence, as an example, we reconstructed  
analytically a new class of scalar-potentials that can lead to the desired 
spectral index and tensor-to-scalar ratio, in agreement with observations. 

Finally, by  imposing other parametrizations for  $P(N)$  we can obtain, 
numerically or 
analytically, other potential forms that lead to the given inflationary 
observables.
Such capabilities reveal the advantages of the use of the scale factor 
potential.

\acknowledgments
This article is supported by COST Action CA15117 "Cosmology and Astrophysics 
Network for Theoretical Advances and Training Action" (CANTATA) of the COST 
(European Cooperation in Science and Technology). This project is partially 
supported by COST Actions CA16104 and CA18108. we thanks to David Vasak for 
additional comments and discussions.

\bibliographystyle{apsrev4-1}
\bibliography{ref}

\end{document}